\documentstyle[12pt]{article}
\catcode`\@=11
\long\def\@makefntext#1{
\protect\noindent \hbox to 3.2pt {\hskip-.9pt
$^{{\ninerm\@thefnmark}}$\hfil}#1\hfill}		%CAN BE USED

 \def\@makefnmark{\hbox to 0pt{$^{\@thefnmark}$\hss}}  %ORIGINAL

\def\ps@myheadings{\let\@mkboth\@gobbletwo
\def\@oddhead{\hbox{}
\rightmark\hfil\ninerm\thepage}
\def\@oddfoot{}\def\@evenhead{\ninerm\thepage\hfil
\leftmark\hbox{}}\def\@evenfoot{}
\def\sectionmark##1{}\def\subsectionmark##1{}}

\newcounter{sectionc}\newcounter{subsectionc}\newcounter{subsubsectionc}
\renewcommand{\section}[1] {\vspace{0.6cm}\addtocounter{sectionc}{1}
\setcounter{subsectionc}{0}\setcounter{subsubsectionc}{0}\noindent
	{\bf\thesectionc. #1}\par\vspace{0.4cm}}
\renewcommand{\subsection}[1] {\vspace{0.6cm}\addtocounter{subsectionc}{1}
	\setcounter{subsubsectionc}{0}\noindent
	{\it\thesectionc.\thesubsectionc. #1}\par\vspace{0.4cm}}
\renewcommand{\subsubsection}[1]
{\vspace{0.6cm}\addtocounter{subsubsectionc}{1}
	\noindent {\rm\thesectionc.\thesubsectionc.\thesubsubsectionc.
	#1}\par\vspace{0.4cm}}

\newcounter{appendixc}
\newcounter{subappendixc}[appendixc]
\newcounter{subsubappendixc}[subappendixc]

\renewcommand{\appendix}[1] {\vspace{0.6cm}
        \refstepcounter{appendixc}
        \setcounter{figure}{0}
        \setcounter{table}{0}
        \setcounter{equation}{0}
        \renewcommand{\thefigure}{\Alph{appendixc}.\arabic{figure}}
        \renewcommand{\thetable}{\Alph{appendixc}.\arabic{table}}
        \renewcommand{\theappendixc}{\Alph{appendixc}}
        \renewcommand{\theequation}{\Alph{appendixc}.\arabic{equation}}
        \noindent{\bf Appendix \theappendixc #1}\par\vspace{0.4cm}}

\def\abstracts#1{{
	\centering{\begin{minipage}{30pc}\tenrm\baselineskip=12pt\noindent
	\centerline{\tenrm ABSTRACT}\vspace{0.3cm}
	\parindent=0pt #1
	\end{minipage}}\par}}

\renewenvironment{thebibliography}[1]
	{\begin{list}{\arabic{enumi}.}
	{\usecounter{enumi}\setlength{\parsep}{0pt}
\setlength{\leftmargin 1.25cm}{\rightmargin 0pt}
	 \setlength{\itemsep}{0pt} \settowidth
	{\labelwidth}{#1.}\sloppy}}{\end{list}}

\newcounter{itemlistc}
\newcounter{romanlistc}
\newcounter{alphlistc}
\newcounter{arabiclistc}

\newcommand{\fcaption}[1]{
        \refstepcounter{figure}
        \setbox\@tempboxa = \hbox{\tenrm Fig.~\thefigure. #1}
        \ifdim \wd\@tempboxa > 6in
           {\begin{center}
        \parbox{6in}{\tenrm\baselineskip=12pt Fig.~\thefigure. #1}
            \end{center}}
        \else
             {\begin{center}
             {\tenrm Fig.~\thefigure. #1}
              \end{center}}
        \fi}

\newcommand{\tcaption}[1]{
        \refstepcounter{table}
        \setbox\@tempboxa = \hbox{\tenrm Table~\thetable. #1}
        \ifdim \wd\@tempboxa > 6in
           {\begin{center}
        \parbox{6in}{\tenrm\baselineskip=12pt Table~\thetable. #1}
            \end{center}}
        \else
             {\begin{center}
             {\tenrm Table~\thetable. #1}
              \end{center}}
        \fi}

\def\@citex[#1]#2{\if@filesw\immediate\write\@auxout
	{\string\citation{#2}}\fi
\def\@citea{}\@cite{\@for\@citeb:=#2\do
	{\@citea\def\@citea{,}\@ifundefined
	{b@\@citeb}{{\bf ?}\@warning
	{Citation `\@citeb' on page \thepage \space undefined}}
	{\csname b@\@citeb\endcsname}}}{#1}}

\newif\if@cghi
\def\cite{\@cghitrue\@ifnextchar [{\@tempswatrue
	\@citex}{\@tempswafalse\@citex[]}}
\def\citelow{\@cghifalse\@ifnextchar [{\@tempswatrue
	\@citex}{\@tempswafalse\@citex[]}}
\def\@cite#1#2{{$\null^{#1}$\if@tempswa\typeout
	{IJCGA warning: optional citation argument
	ignored: `#2'} \fi}}

\def\fnt#1#2{\footnotetext{\kern-.3em
	{$^{\mbox{\sevenrm #1}}$}{#2}}}

 1
 1
 1

\font\tenbf=cmbx10
\font\tenrm=cmr10
\font\tenit=cmti10

\font\ninerm=cmr9

\begin{document}
\begin{flushright}
TTP 95-04\\
hep-ph/9502357
\end{flushright}
\centerline{\tenbf RECENT RESULTS ON QCD CORRECTIONS}
\baselineskip=22pt
\centerline{\tenbf TO SEMILEPTONIC $b$-DECAYS\footnote{Talk given at
the WE-Heraeus Seminar  ``Heavy Quark Physics'', Dec.~1994,
Bad Honnef, Germany}}
\baselineskip=16pt
\vspace{0.8cm}
\centerline{\tenrm ANDRZEJ CZARNECKI and MAREK JE\.ZABEK\footnote{
Permanent address: Institute of Nuclear Physics, Kawiory 26a, PL-30055
Cracow, Poland}}
\baselineskip=13pt
\centerline{\tenit Institut f\"ur Theoretische Teilchenphysik,
Universit\"at Karlsruhe,}
\baselineskip=12pt
\centerline{\tenit D-76128 Karlsruhe, Germany}
\vspace{0.9cm}
\abstracts{We summarize recent results on QCD corrections to various
observables in semileptonic $b$ quark decays.  For massless leptons in
the final state we present effects of such corrections on triple
differential distribution of leptons which are important in studies of
polarized $b$ quark decays.  Analogous formulas for distributions of
neutrinos are applicable in decays of polarized $c$ quarks. In the
case of decays with a $\tau$ lepton in the final state mass effect of
$\tau$ has to be included.  In this case we concentrate on corrections
to the total decay width.}

\vfil
\rm\baselineskip=14pt
\section{Introduction}
Studies of the $b$ are one of the most promising directions in the
high energy physics in the '90s\cite{viv,alb}.  Wealth of recent results from
experiments at LEP and CESR, together with older data from ARGUS at
DESY, already determine some properties of the
$b$ sector very well.  In the relatively near future new
experiments HERA-B at DESY, LHC-B at CERN, and $B$-factories at SLAC
and at KEK, will be able to study $b$ physics with high precision.
Therefore it is timely to prepare precise theoretical predictions for
the properties of decays of the $b$ quark.

In this paper we summarize our recent results concerning QCD
corrections in semileptonic $b$ decays. Section 2 deals with
energy-angular distributions of massless leptons and section 3
with the total width of the decay $b\to c \tau\bar\nu_\tau$.

\section{Polarized bottom and charm quarks}

Polarization studies for heavy flavors at LEP are a new
interesting field of potentially fundamental significance,
see Refs.3 and 4 for recent reviews.
According to the Standard Model
$Z^0\to b\bar b$ and
$Z^0\to c\bar c$ decays
can be viewed as sources of highly polarized heavy quarks.
The degree of longitudinal polarization is fairly large,
amounting to $\langle P_b\rangle = -0.94$ for $b$
and $\langle P_c\rangle = -0.68$ for $c$ quarks\cite{Kuehn1}.
The polarizations depend weakly on the production angle.
QCD corrections to Born result are about 3\%\cite{KPT}.
The real drawback is that due to hadronization the net longitudinal
polarization of the decaying  $b$ and $c$ quarks is drastically
decreased. In particular these $b$ quarks become depolarized
which are bound in $B$ mesons
both produced directly and from $B^*\to B\gamma$ transitions.
The signal is therefore significantly reduced.
Only those $b$'s (a few percent) which fragment directly
into $\Lambda_b$  baryons retain information on the original
polarization\cite{Bjo}.
Polarization transfer from a heavy
quark $Q$ to the corresponding $\Lambda_Q$ baryon is 100\%
\cite{CKPS} at least in the limit $m_Q\rightarrow\infty$.
Thus, a large net polarization is expected for heavy quarks
in samples enriched with these heavy baryons.

It has been proposed long ago\cite{zerwas}
that distributions of charged leptons from semileptonic
decays of beautiful hadrons can be used in polarization
studies for $b$ quarks.
Some advantages of neutrino distributions have been
also pointed out\cite{CJKK,BR,dittmar}.
Recently there has been considerable progress in the theory of
the inclusive semileptonic decays of heavy flavor hadrons.
It has been shown that in the leading order of an expansion
in inverse powers of heavy quark mass $1/m_Q$ the
spectra for hadrons coincide with those for the decays of
free heavy quarks\cite{CGG} and there are no $\Lambda_{QCD}/m_Q$
corrections to this result away from the energy
endpoint.
$\Lambda^2_{QCD}/m_Q^2$ corrections have been
calculated in Refs.14,15 for $B$ mesons and
in ref.15 for polarized $\Lambda_b$ baryons.
For some decays the results are similar to those of the well-known
$ACCMM$ model\cite{altar}.
The corrections to charm decays are larger than for bottom
and convergence of
$1/m_Q$  expansion is poorer\cite{Shifman}.
Perturbative first order QCD corrections contribute 10-20\%
to the semileptonic decays and for bottom are larger
than the nonperturbative ones.

In our recent article\cite{CJ} compact analytic
formulae have been obtained
for the distributions of the charged lepton and the neutrino.
These formulae  agree with the results of our earlier
calculations\cite{CJK,CJKK}
for the joint angular and energy distribution
of the charged lepton in top, charm and bottom quark
decays.
The QCD corrected triple differential distribution
of the charged lepton
for the semileptonic decay  of the polarized quark
with the weak isospin $I_3=\pm 1/2$
can be written in the following way\cite{CJ}:
\begin{eqnarray}
{{\rm d}\Gamma^{\pm} \over {\rm d}x\,{\rm d}y\,{\rm d}\cos\theta  }
&& \sim\qquad \left[\,
{\rm F}^\pm_0(x,y) + S\cos\theta\,{\rm J}^\pm_0(x,y)\,
\right]
\nonumber\\
&& -\; {2\alpha_s\over3\pi}\;
  \left[\,
  {\rm F}^\pm_1(x,y) + S\cos\theta\,{\rm J}^\pm_1(x,y)\,
  \right]
\nonumber\\
\label{eq:1}
\end{eqnarray}
In the rest frame of the decaying heavy quark
$\theta$ denotes the angle between the
polarization vector $\vec s$ of the heavy quark and the
direction of the charged lepton,
$S=|\,\vec s\,|$,
$x= 2Q\ell/Q^2$ and $y= 2\ell\nu/Q^2$ where
$Q$, $\ell$ and $\nu$ denote the four-momenta of the decaying
quark, charged lepton and neutrino. Eq.(\ref{eq:1}) describes also
the triple differential
distribution of the neutrino for  $I_3=\mp 1/2$. In this case,
however, $x= 2Q\nu/Q^2$  and
$\theta$ denotes the angle between
$\vec s$ and the three-momentum of the neutrino.
The functions ${\rm F}^\pm_0(x,y)$ and ${\rm J}^\pm_0(x,y)$
corresponding to Born approximation read:
\begin{eqnarray}
{\rm F}^+_0(x,y) &=& x (x_m-x)
\label{eq:5}\\
{\rm J}^+_0(x,y) &=& {\rm F}^+_0(x,y)
\label{eq:6}\\
{\rm F}^-_0(x,y) &=& (x-y) (x_m-x+y)
\label{eq:7}\\
{\rm J}^-_0(x,y) &=& (x-y) (x_m-x+y-2y/x)
\label{eq:8}
\end{eqnarray}
where $x_m=1-\epsilon^2$, $\epsilon^2= q^2/Q^2$ and $q$
denotes the four-momentum of the quark originating from
the decay.
The functions ${\rm F}^\pm_1(x,y)$ and
${\rm J}^+_1(x,y)$ correspond to
the first order QCD corrections and
are given in Ref.18.

Non-trivial cross checks are fulfilled by the polarization
independent parts of the distributions (\ref{eq:1}):
\begin{itemize}
\item
the distributions
${\rm d}\Gamma^{\pm} / {\rm d}x\,{\rm d}y$
agree with the results
for unpolarized decays which were
obtained by Je\.zabek and K\"uhn\cite{JK}.
The present formulae are simpler.
\item
in the four-fermion (Fermi) limit integration over $y$
can be performed numerically. The resulting distributions
${\rm d}\Gamma^{\pm} / {\rm d}x$
also agree with those of Ref.20.
Recently the results of Ref.20
have been confirmed\cite{CCM1}. Thus an old conflict
with other calculations\cite{CCM0} is solved and
the agreement with Ref.20 can be considered as a
non-trivial cross check. Moreover, the
analytic result\cite{JK} for
${\rm d}\Gamma^{+} / {\rm d}x$
and $\epsilon=0$
has been also confirmed\cite{GWF}.
\item
$${{\rm d}\Gamma^{+} / {\rm d}y} =
{{\rm d}\Gamma^{-} / {\rm d}y}$$
and the analytic formula for this distribution
exists\cite{JK0} which at the same time describes the lifetime
of the top quark as a function of its mass.
This formula has been confirmed by a few groups, c.f. Ref.25
and references therein.
\item
in the four-fermion limit the result for the total rate $\Gamma$
derived from
eq.(\ref{eq:1})
agrees with the results of Ref.26 and
the analytical formula of Ref.27.
\end{itemize}

\section{QCD corrections to $b\to c \tau\bar\nu_\tau$}
The semileptonic decay of $b$ quark in the case of massive lepton in
the final state is a particularly interesting process.  In contrast to
the case of light leptons the decay rate is sensitive to the
coupling of the scalar component of the $W$ boson.  It can also be
strongly affected by the charged Higgs boson predicted by many
extensions of the standard model, e.g. by the minimal supersymmetric
standard model.  Recently the branching ratio has  been
measured: Br$(b\to \tau\nu X)$ = $(2.75\pm 0.3 \pm 0.3)\%$\cite{viv}.

The perturbative QCD corrections to the inclusive decay rate have
been evaluated numerically\cite{FLNN}. The bound stated
corrections have also been computed using Heavy Quark Effective
Theory\cite{FLNN,Koyrakh,BKPS}.  Recently we have presented the
analytical formula for the QCD corrections in the form of a one
dimensional intergral over the invariant mass squared ${\rm w}^2$ of
the leptons\cite{cjk94a}:
\begin{equation}
\Gamma(b\to c\tau\bar\nu_\tau)\:=\:
\int^{(1-\sqrt{\rho})^2}_\eta
{{\rm d}\Gamma\over {\rm d}t}\,{\rm d}t
\label{eq:Gamtot}
\end{equation}
with the
differential rate
\begin{equation}
{{\rm d}\Gamma\over {\rm d}t} = \Gamma_{bc}\,
\left( 1-{\eta \over t}\right)^2 \,  \left\{
\left( 1+{\eta \over 2t}\right)
\left[ {\cal F}_0(t) - {2\alpha_s\over 3 \pi} {\cal F}_1(t)\right]
+ {3\eta \over 2t}
\left[ {\cal F}_0^s(t) - {2\alpha_s\over 3 \pi} {\cal F}_1^s(t)\right]
\right\}
\label{main}
\end{equation}
where
\begin{equation}
\Gamma_{bc} = {G_F^2 m_b^5 |V_{cb}|^2\over 192 \pi^3}
\end{equation}
and we have used the following dimensionless variables
$$
\textstyle{
\rho= {m_c^2/ m_b^2}\qquad\quad
\eta = {m_\tau^2/ m_b^2}\qquad\quad
{\rm and} \qquad\quad
t = {{\rm w}^2/m_b^2  }  }
$$
and the explicit formulas for the functions ${\cal F}_i$ are given,
together with a detailed derivation, in our
paper\cite{cjk94a}.  The integrated  decay rate can be rewriten as
\begin{equation}
\Gamma(b\to c\tau\bar\nu_\tau)\:=\: \Gamma^{(0)}\left[
1-{2\alpha_s\over 3\pi}
F(m_b,m_c,m_\tau)\right]
\end{equation}
In Fig.~1 we present the dependence of the relative correction
$F(m_b,m_c,m_\tau)$ on $m_b$ for two different values of
$m_b-m_c$.  We also plot an analogous function $F(m_b,m_c,0)$
relevant for the decays with electrons or muons in the final state.
One can see that the relative corrections are smaller for a heavier
lepton and for a heavier quark  in the final state.

An important feature of formula \ref{main} is that it gives the
QCD correction to the decay rate differentiated with respect to the
square of the invariant mass of the lepton system.  The perturbation
theory is more reliable in the region of smaller ${\rm w}^2$, so it is
useful to have the correction given as a function of ${\rm w}^2$.  We
rewrite eq.~\ref{main} as
\begin{equation}
{{\rm d}\Gamma\over {\rm d}t} ={{\rm d}\Gamma^{(0)}\over {\rm d}t}
\left[ 1-{2\alpha_s\over 3\pi}G(t)\right].
\end{equation}
The shape of the normalized Born distribution is shown
in Fig.~2a. The function $G(t)$ is plotted in
Fig.~2b for $\tau$ as well as for massless leptons.

Further theoretical study of the decay $b\to c \tau\bar\nu_\tau$ is
needed.  Especially important is the determination of QCD corrections
to the energy spectrum of the $\tau$.  This is important for the
current measurements of the branching ratio, since these are based on
an assumed distribution of the missing energy.  It will become even
more important in the future measurements of the leptonic decay
$B\to \tau\bar\nu_\tau$, where $\tau$'s from the channel
 $b\to c \tau\bar\nu_\tau$ are the major background.  The
leptonic decay is going to be the source of determination of the $B$
decay constant $f_B$, crucial in the efforts to determine the weak
mixing angles and overconstrain the
unitarity triangle.

The QCD corrections to the semileptonic decay of $b$ into $\tau$
are also applicable to the decay $b\to c\bar c s$.
\begin{equation}
\Gamma(b\to c\bar cs) =
\Gamma^{(bare)}_{c\bar cs}\, \left[
1 + {\textstyle {\alpha_s\over\pi}}\, \left(
\delta_{bc} + \delta_{\bar cs} + \delta_{penguin}\, \right)
\,\right]
\end{equation}
where $\Gamma^{(bare)}_{c\bar cs}$ is the rate without any QCD
corrections, $\delta_{bc}$ arises from gluon exchange on the
$bc$ line, $\delta_{\bar cs}$ is due to gluon interactions
within the ${\bar cs}$ loop, and $\delta_{penguin}$
is due to effects of the penguin type; cf.~Ref.~32
where $\delta_{\bar cs}$ is calculated as an integral over the
invariant mass of the ${\bar cs}$ system.
Our calculation provides the missing part of the correction
$\delta_{bc}$ (see also Ref.~33).
We plot $\delta_{bc}$ in Fig.~3 for four different
values of $m_b-m_c$.

%\pagebreak[4]

\section{Acknowledgments}We would like to thank J\"urgen K\"orner and
Hans K\"uhn for collaboration on research reported in this article.
A.C. thanks Vivek Sharma for helpful discussion, and the organizers
of the {WE-Heraeus-Seminar} at Bad Honnef for hospitality. This work is
partly supported by EEC network CIPDCT 94 0016 and by KBN grant
2P30207607.

\section{List of figures}
\begin{enumerate}
\item{Relative QCD correction to the total decay width as a
function of mass of the $b$ quark for $m_b-m_c$ = 3.414 GeV and 3.317
GeV:
for  a massless lepton (solid and dashed lines) and for the $\tau$
lepton (dotted and dash-dotted lines).}
%\label{fig:ftau}
\item{(a) Normalized Born differential decay rates for massless
leptons (solid) and for the $\tau$ lepton (dashed);
(b)  QCD correction $G(t)$  to the differential decay width
for  a massless lepton (solid) and for  $\tau$ (dotted).}
%\label{fig:tdep}
\item{QCD correction $\delta_{bc}$ for the decay $b\to c\bar c s$
as a function of $m_b$, for various values of $m_b-m_c$: 3.2 GeV
(solid),  3.3 GeV (dashed), 3.4 GeV (dotted), and  3.5 GeV (dash-dotted).}
%\label{fig:charm}
\end{enumerate}

\end{document}